\documentclass[9pt,shortpaper,twoside,web]{ieeecolor}
\usepackage{generic}
\usepackage{cite}
\usepackage{amsmath,amssymb,amsfonts}
\usepackage{algorithmic}
\usepackage{graphicx}
\usepackage{textcomp}
\usepackage{hyperref}
\usepackage{makecell}
\usepackage{bbding}
\usepackage{threeparttable}
\def\BibTeX{{\rm B\kern-.05em{\sc i\kern-.025em b}\kern-.08em
    T\kern-.1667em\lower.7ex\hbox{E}\kern-.125emX}}
\begin{document}
\title{Breast Cancer Immunohistochemical Image Generation: \\a Benchmark Dataset and Challenge Review}
\author{Chuang Zh, Shengjie Liu, Zekuan Yu, Feng Xu, Arpit Aggarwal, Germán Corredor, Anant Madabhushi, Qixun Qu, Hongwei Fan, Fangda Li, Yueheng Li, Xianchao Guan, Yongbing Zhang, Vivek Kumar Singh, Farhan Akram, Md. Mostafa Kamal Sarker, Zhongyue Shi and Mulan Jin
\thanks{The first two authors contributed equally to this work.}
\thanks{This work was supported in part by National Key R\&D Program of China (2021ZD0109800), and in part by the National Natural Science Foundation of China (81972248).}
\thanks{C. Zhu and S.J. Liu are with the School of Artificial Intelligence, Beijing University of Posts and Telecommunications, Beijing 100876, China.}
\thanks{Z.K. Yu is with Academy for Engineering and Technology, Fudan University, Shanghai, China.}
\thanks{F Xu, Z.Y. Shi and M.L. Jin are with Beijing Chaoyang Hospital, Capital Medical University
Beijing, China. }
\thanks{Corresponding authors:  Zekuan Yu and Feng Xu (yzk@fudan.edu.cn; drxufeng@mail.ccmu.edu.cn)}
\thanks{Ethics committee/IRB of Beijing Chao-Yang Hospital, Capital Medical University gave ethical approval for this work.}
}
\maketitle

\begin{abstract}
For invasive breast cancer, immunohistochemical (IHC) techniques are often used to detect the expression level of human epidermal growth factor receptor-2 (HER2) in breast tissue to formulate a precise treatment plan. From the perspective of saving manpower, material and time costs, directly generating IHC-stained images from Hematoxylin and Eosin (H\&E) stained images is a valuable research direction. Therefore, we held the breast cancer immunohistochemical image generation challenge, aiming to explore novel ideas of deep learning technology in pathological image generation and promote research in this field. The challenge provided registered H\&E and IHC-stained image pairs, and participants were required to use these images to train a model that can directly generate IHC-stained images from corresponding H\&E-stained images. We selected and reviewed the five highest-ranking methods based on their PSNR and SSIM metrics, while also providing overviews of the corresponding pipelines and implementations. 
In this paper, we further analyze the current limitations in the field of breast cancer immunohistochemical image generation and forecast the future development of this field. We hope that the released dataset and the challenge will inspire more scholars to jointly study higher-quality IHC-stained image generation.
\end{abstract}

\begin{IEEEkeywords}
Breast cancer, pathology image dataset, immunohistochemical image generation, image-to-image translation, grand challenge.
\end{IEEEkeywords}

\section{Introduction}
\label{sec:introduction}
According to data \cite{sung2021global} released by the International Agency for Research on Cancer (IARC), female breast cancer has surpassed lung cancer as the most commonly diagnosed cancer in 2020, with an estimated 2.3 million new cases. Early determination of the type and stage of breast cancer is crucial to the formulation of treatment plans and the prognosis of patients.  

\begin{figure}[htbp]
\centering
    \begin{minipage}[t]{0.48\linewidth}
        \centering
        \includegraphics[width=\textwidth]{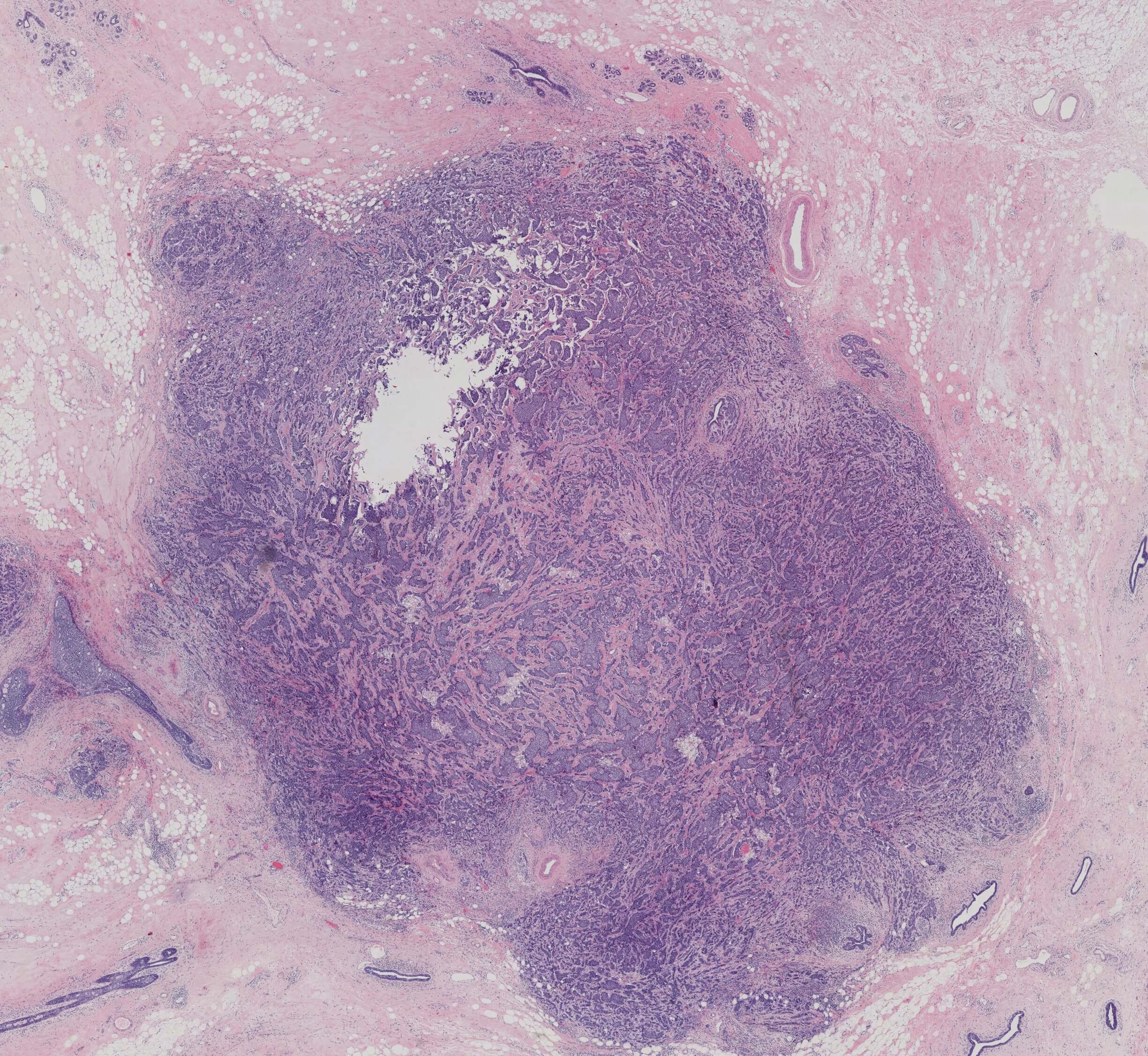}
        \centerline{\footnotesize(a) An example of H\&E slice.}
    \end{minipage}%
    \hspace{2mm}
    \vspace{3mm}
    \begin{minipage}[t]{0.48\linewidth}
        \centering
        \includegraphics[width=\textwidth]{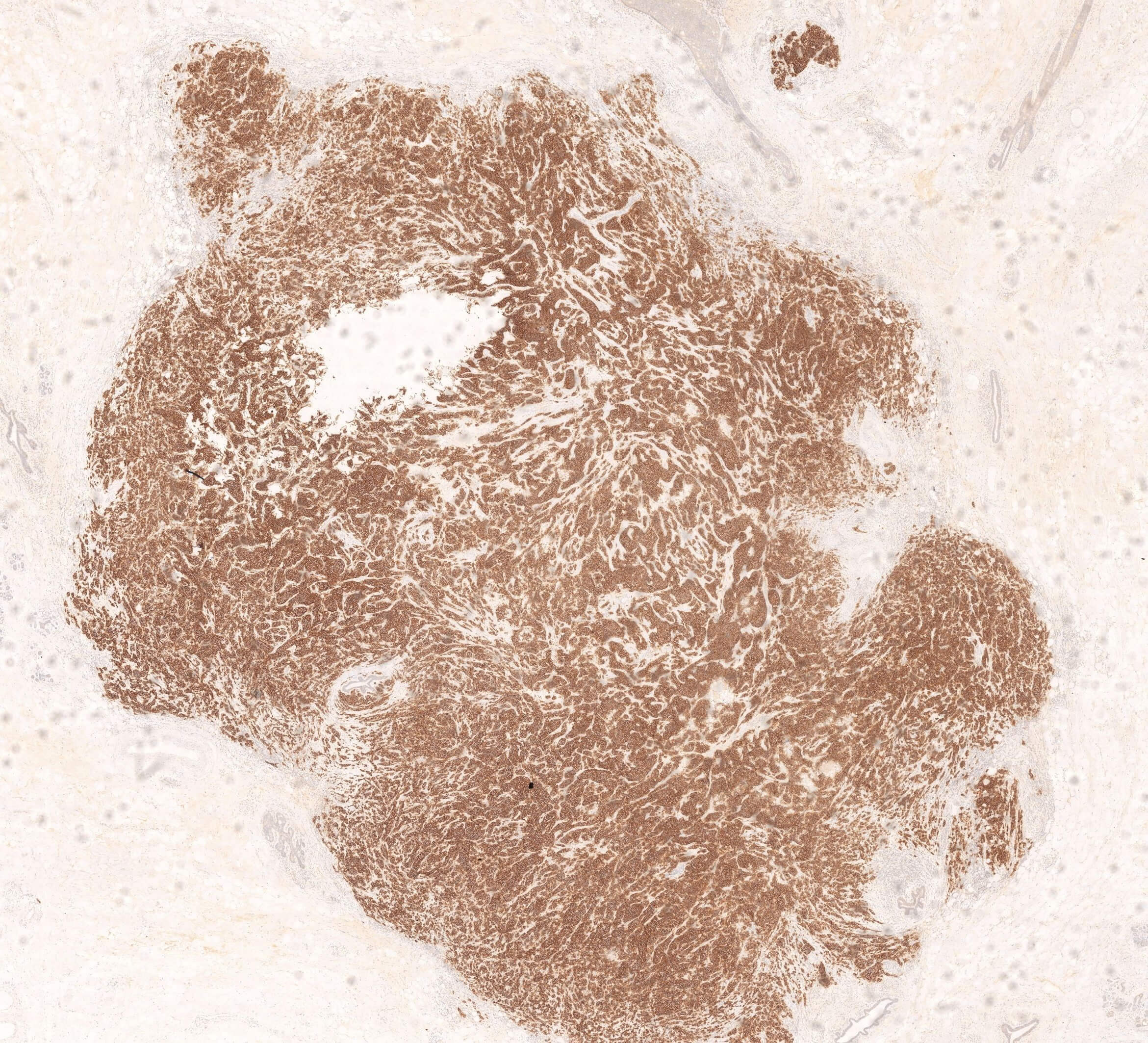}
        \centerline{\footnotesize(b) An example of IHC-stained slice.}
    \end{minipage}
    \caption{Visualization of an H\&E-stained slice and the corresponding immunohistochemical (IHC) stained slice.}
    \label{WSIexamples}
\end{figure}

The current diagnosis of breast cancer is based on the pathological tissue stained with Hematoxylin and Eosin (H\&E) as the gold standard. Surgeons take a piece of tissue from the lesion area of the patient, which undergoes a series of procedures including slice preparation and staining, to finally make a pathological slide available for observation. The pathologist then observes the slice under a microscope and provides a diagnosis. An H\&E-stained slice is shown in Fig.~\ref{WSIexamples}(a)).

For patients diagnosed with breast cancer, specific protein testing is often required to further evaluate the tumor. For example, the state (positive or negative) of human epidermal growth factor receptor-2 (HER2), needs to be identified for breast cancer, as the HER2 state is a helpful marker for therapy decision making \cite{iqbal2014human}.

If a patient tests positive for HER2, doctors will administer targeted drug therapy. Timely targeted therapy can increase the survival chance of HER2-positive patients to a level similar to those of HER2-negative patients. Experts recommend that all patients diagnosed with invasive breast cancer undergo HER2 testing to significantly improve the treatment recommendations and decisions \cite{yamauchi2008her2}.

\begin{figure}[htbp]
\centering
    \begin{minipage}[t]{0.47\linewidth}
        \centering
        \includegraphics[width=\textwidth]{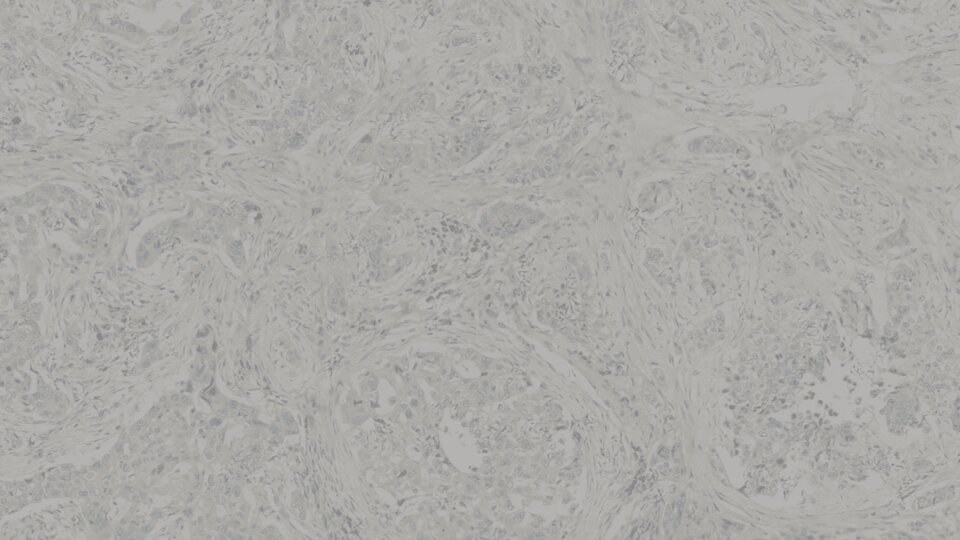}
        \centerline{\footnotesize{(a) IHC 0}}
    \end{minipage}
    \hspace{2mm}
    \vspace{3mm}
    \begin{minipage}[t]{0.47\linewidth}
        \centering
        \includegraphics[width=\textwidth]{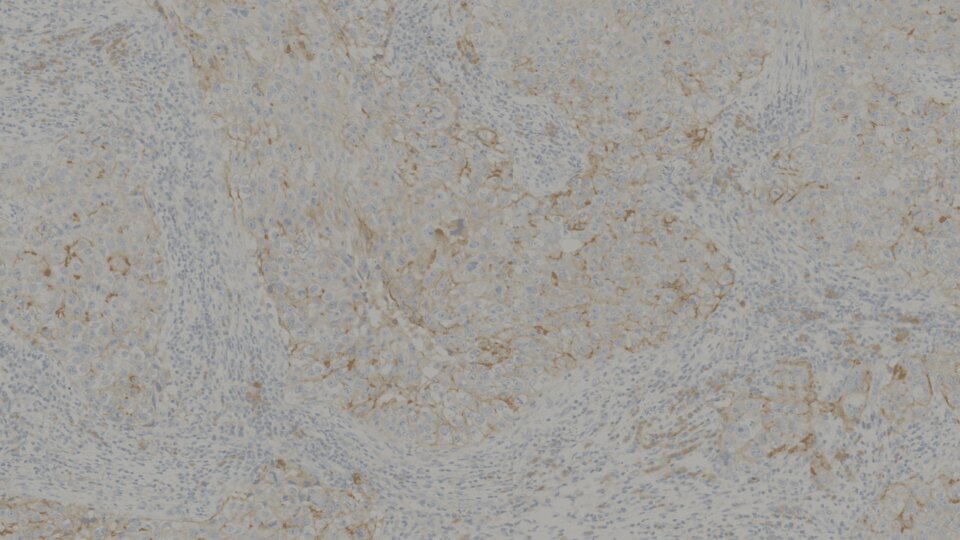}
        \centerline{\footnotesize{(b) IHC 1+}}
    \end{minipage}
    \vspace{3mm}
    \begin{minipage}[t]{0.47\linewidth}
        \centering
        \includegraphics[width=\textwidth]{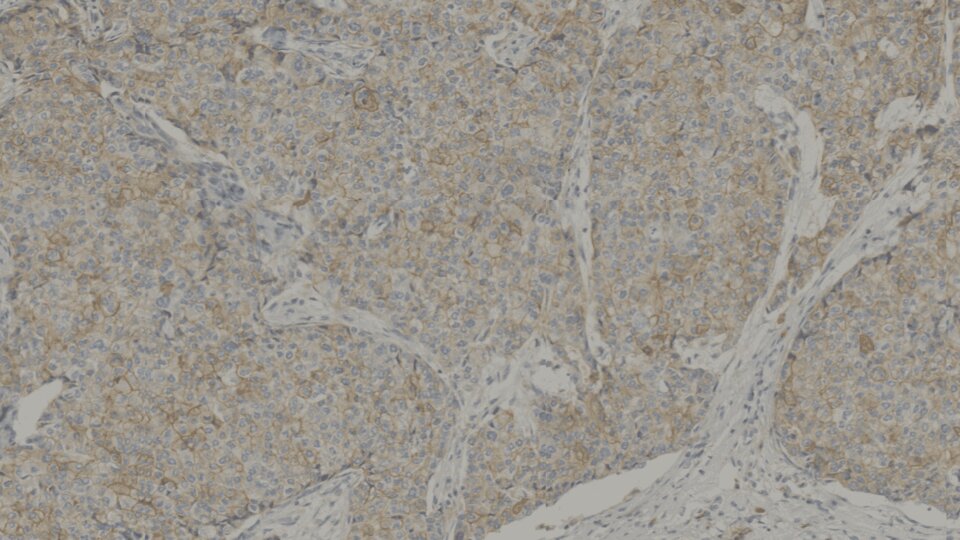}
        \centerline{\footnotesize{(c) IHC 2+}}
    \end{minipage}
    \hspace{2mm}
    \begin{minipage}[t]{0.47\linewidth}
        \centering
        \includegraphics[width=\textwidth]{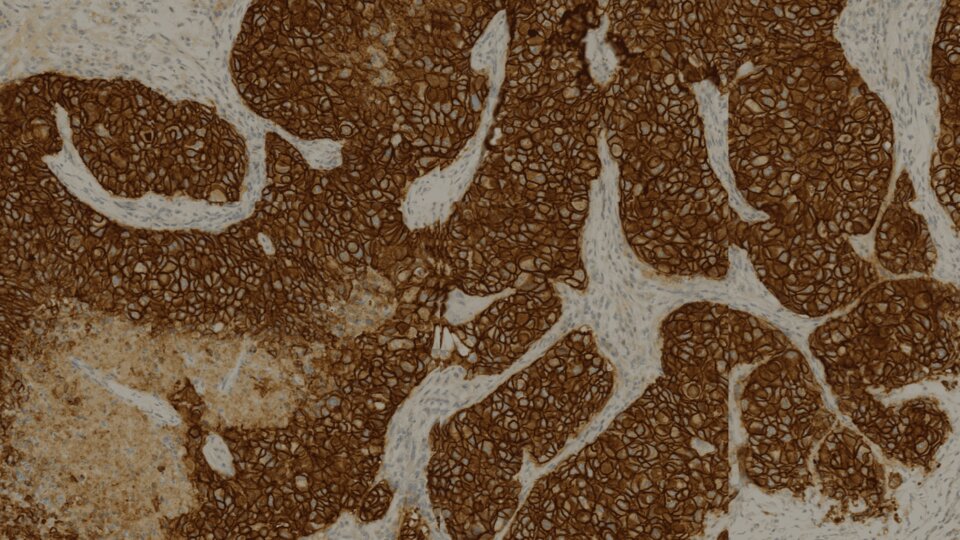}
        \centerline{\footnotesize{(d) IHC 3+}}
    \end{minipage}
    \caption{Visualization of different HER2 expression levels. Generally, the cell membrane is stained darker with increasing HER2 expression levels.}
    \label{HER2expressions}
\end{figure}

The normal method for evaluating HER2 expression levels is to interpret the pathological images stained by immunohistochemical (IHC) technique. Specifically, an additional tissue section is taken from the patient's pathological tissue for IHC staining (an IHC-stained slice is shown in Fig.~\ref{WSIexamples}(b)), and pathologists determine the level of HER2 expression based on the staining pattern of the cell membrane in the section. According to clinical oncology practical guideline of American \cite{wolff2018human}, the interpretation rules for immunohistochemical images of breast cancer are as follows: \textit{IHC 0, no staining is observed or membrane staining that is incomplete and is faint/barely perceptible and in $\leq 10\%$ of tumor cells} (Fig.~\ref{HER2expressions}(a)); \textit{IHC 1+, incomplete membrane staining that is faint/barely perceptible and in \textgreater 10\% of tumor cells} (Fig.~\ref{HER2expressions}(b)); \textit{IHC 2+, weak to moderate complete membrane staining observed in \textgreater 10\% of tumor cells} (Fig.~\ref{HER2expressions}(c)); \textit{IHC 3+, circumferential membrane staining that is complete, intense, and in \textgreater 10\% of tumor cells} (Fig.~\ref{HER2expressions}(d)).

\begin{figure}[!t]
\centering
\includegraphics[height = 230pt, width=260pt]{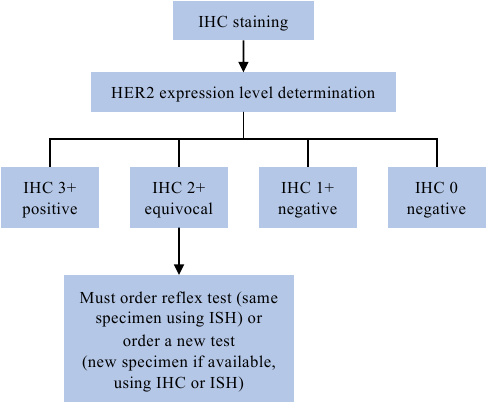}
\caption{Algorithm for evaluation of human epidermal growth factor receptor 2 (HER2) protein expression by immunohistochemistry (IHC) assay of the invasive component of a breast cancer specimen.}
\label{IHC_class}
\end{figure}

The evaluation of HER2 expression is critical to the formulation of follow-up treatment plans for breast cancer. However, it is expensive to conduct HER2 evaluation through the preparation of an IHC-stained slice. Moreover, a single IHC-stained section may not be able to comprehensively assess the level of HER2 expression in tumor tissue. For instance, in Fig.~\ref{IHC_class}, if the HER2 expression level is  diagnosed as 2+, obtaining a new tissue sample for IHC staining may be necessary. This additional preparation requirement for IHC-stained slices further increases the costs associated with labor and materials.

The question that arises is whether it is possible to synthesize an IHC-stained image from an H\&E-stained image, and thus avoid the expensive IHC staining. With the advancement of deep learning, many intelligent applications in the medical field have emerged, such as tumor cell classification \cite{zhang2017deeppap,xie2022deep,pattarone2021learning}, tumor segmentation \cite{havaei2017brain,wang2021transbts,jiang2022swinbts}, and pathology image staining normalization \cite{shaban2019staingan,lee2022stain,perez2022staincut}, etc. The powerful capabilities demonstrated by deep learning in the above-mentioned fields make us hope that it can also have the ability to directly generate IHC-stained pathological images.
% \added{Given this, there is also potential for deep learning models to generate IHC-stained pathology images.}
Successful exploration of this technology would help save considerable human and financial costs associated with IHC-stained slice preparation. At the same time, to mitigate potential inaccurate HER2 detection stemming from tumor heterogeneity \cite{turashvili2017tumor}, IHC-stained image generation could easily be performed on multiple H\&E-stained tumor tissue sections from invasive breast cancer patients. 
% If successful, this would enable us to directly evaluate HER2 expression based on the synthesized IHC-stained images. 
Besides, this study will also help us understand and interpret what information in H\&E-stained images is relevant to HER2 assessment in the future. 

The dataset is a crucial factor in studying H\&E to IHC-stained image translation, and we have successfully collected and established the Breast Cancer Immunohistochemical (BCI) \footnote{\href{https://bupt-ai-cz.github.io/BCI_for_GrandChallenge}{https://bupt-ai-cz.github.io/BCI\_for\_GrandChallenge}}, a paired H\&E to IHC-stained image translation dataset. In this dataset, H\&E-stained images and IHC-stained images are already achieved structure-level alignment.

The dataset provides a foundation for the research of the IHC-stained image generation algorithms. Based on the BCI dataset, we hosted a challenge\footnote{\href{https://bci.grand-challenge.org}{https://bci.grand-challenge.org}} for the generation of breast cancer IHC-stained images, in which participants were required to train an IHC-stained image generation algorithm and submit the generated IHC-stained images. The challenge attracted over 500 registrations and received a total of 75 submissions.

\section{Related Work}

\subsection{Computer-Aided Diagnosis of Pathology}
 Deep learning has been widely used in many computer vision tasks such as image classification \cite{he2016deep}, semantic segmentation \cite{ronneberger2015u}, and object detection \cite{ren2015faster}. The extension of the above technologies in the field of pathology has also become a hot research topic. The current applications of deep learning in pathological image analysis include tumor detection and classification, tumor segmentation, cell detection and counting, etc.

 The pioneering work \cite{janowczyk2016deep} gave a series of benchmarks for pathological image-based detection, segmentation and recognition based on Convolutional Neural Networks (CNN). These benchmarks include nuclei segmentation, epithelium segmentation, tubule segmentation, Invasive Ductal Carcinoma (IDC) segmentation, lymphocyte detection, mitosis detection, and lymphoma sub‑type classification. Inspired by the above work and driven by demand, a large number of researches on classification \cite{shao2021transmil,zou2022breast} and segmentation \cite{song2020clinically,wang2021medical,greenwald2022whole} of pathological images  based on deep learning have emerged. Work \cite{shao2021transmil} designed a transformer-based Multiple Instance Learning (MIL) framework, which can effectively deal with unbalanced/balanced and binary/multiple WSI classification. Work \cite{zou2022breast} introduced a novel Attention High-order deep Network (AHoNet) by simultaneously embedding attention mechanism and high-order statistical representation into a residual convolutional network and this network can capture more discriminative deep features for breast cancer pathological images. Regarding the detection and segmentation of specific pathological tissue regions, works \cite{song2020clinically} and \cite{wang2021medical} used semantic segmentation models (i.e. DeepLabv3 \cite{chen2017rethinking}, DeepLabv3+ \cite{chen2018encoder}) to generate candidate cancer regions in pathological images for reference by pathologists. Work \cite{greenwald2022whole} introduced TissueNet, an extensively annotated tissue image dataset designed for training cell segmentation models. This dataset was employed to train Mesmer, a segmentation model that outperforms previous algorithms in terms of accuracy. Furthermore, some researchers used multi-task learning to analyze pathological images \cite{graham2023one} to realize the segmentation and classification tasks in one model. These studies have greatly promoted the development of computer-aided diagnosis. In clinical practice, some  mature algorithms have been deployed to the front line, and these algorithms are able to automate repetitive and time-consuming tasks, playing a huge role in reducing the clinical workload of pathologists.
 
\subsection{Classification of HER2 Expression Levels}

To explore alternatives to HER2 expression level classification without relying on IHC-stained images, some scholars have begun to employ deep learning-based methods to predict HER2 expression levels based on images from other modalities, thereby eliminating the IHC staining step. Xu et al. \cite{xu2022predicting} proposed a DenseNet-based deep learning model using ultrasound images as input to predict HER2 expression and the performance significantly exceeded the traditional texture analysis based on the radiomics model. La Barbera et al. \cite{la2020detection} proposed a pipeline that mimics clinician diagnosis: they first employed a cascade of deep neural network classifier for breast cancer screening and then detected the presence of HER2 via MIL. In addition, there are some other studies \cite{anand2020deep,farahmand2022deep} predict the expression level of HER2 based on H\&E-stained images. Work \cite{anand2020deep} proposed a multi-stage image classification pipeline to realize the classification of breast cancer tumors based on H\&E-stained images. Work \cite{farahmand2022deep} utilized an Inceptionv3 \cite{szegedy2016rethinking} architecture to predict HER2 status in breast cancer and trastuzumab treatment in HER2-positive samples.

The above studies that predict HER2 expression levels based on H\&E-stained images demonstrate that such images contain some information regarding HER2 expression levels. However, the output of these classification algorithms is HER2 positive probability, and the classification model is a black box, which limits the interpretability of the model's HER2 expression level predictions.

\subsection{Image-to-Image Translation}
Image translation algorithms can be divided into supervised image translation algorithms and unsupervised image translation algorithms according to the form of supervision. For supervised image translation models, pixel-level aligned image pairs are required for supervision during the training phase. Some pioneering works \cite{isola2017image, wang2018high} have implemented supervised translation of natural images. Image translation algorithms have significant application value in the field of pathology. Recently, work \cite{Liu_2022_CVPR} proposed the PyramidPix2pix model to constrain the generated images on multiple scales and achieved state-of-the-art (SOTA) on the pathology image translation dataset. The impressive work \cite{ghahremani2022deep} employed the DeepLIIF framework, which is capable of converting IHC images into more informative and higher-cost Multiplex Immunofluorescence (mpIF) images. Different from the above supervised image translation algorithms, unsupervised image translation models \cite{zhu2017unpaired,liu2017unsupervised,huang2018multimodal,lee2018diverse} are trained with unaligned image pairs and this type of algorithms realize changing the image style mainly by adversarial learning. In the field of medical image analysis, unsupervised image translation algorithms are primarily employed for the staining normalization of pathological images \cite{cho2017neural,shaban2019staingan,cai2019stain} and data augmentation \cite{han2019synthesizing,gupta2019generative,han2019combining,mahmood2019adversarial,liu2020isocitrate,stacke2020measuring}. 

In this study, we aim to leverage image-to-image translation technology to generate IHC-stained images from corresponding H\&E-stained images. This approach would enable us to better visualize the expression of HER2 in tumor tissues, rather than directly predicting the HER2 status.

\section{Dataset Construction}

As a key factor to improve the performance of deep learning models, many open-source datasets have been widely applied in computer vision tasks, such as ImageNet \cite{deng2009imagenet} and MNIST \cite{lecun1998gradient} for image classification; COCO \cite{lin2014microsoft}, Cityscapes \cite{cordts2016cityscapes}, and KITTI \cite{geiger2012we} for semantic segmentation and object detection.

However, there are currently no publicly available datasets for conducting research on generating breast cancer IHC-stained images. Study \cite{van2021deep} highlights that a major challenge in the current field of Computational Pathology (CPATH) is the lack of publicly available datasets that truly represent clinical practice. Therefore, we introduced the BCI dataset. This dataset serves as a valuable resource for investigating the conversion of H\&E-stained breast cancer tissue images into IHC-stained images. The forthcoming section will comprehensively outline the construction methodology employed for the development of the BCI dataset.
% Therefore, we introduced the BCI dataset, this dataset can be utilized to study translating H\&E-stained images of breast cancer tissue into IHC-stained images, offering a new approach for HER2 detection or other downstream tasks based on IHC-stained images. The following section will provide a detailed account of the construction process of the BCI dataset.

\begin{figure*}[!t]
\centering
\includegraphics[width=1.0\linewidth]{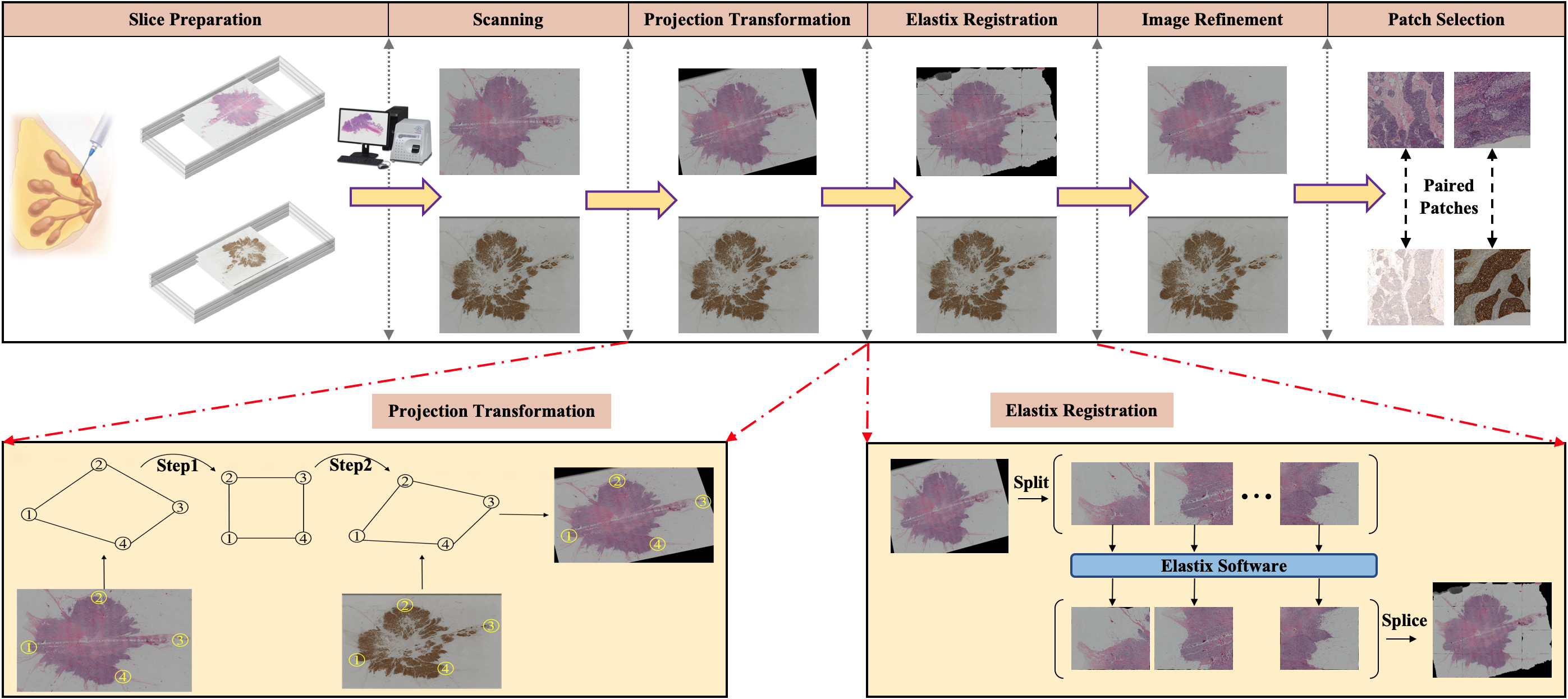}
\caption{The process of constructing the BCI dataset. First, we prepared H\&E-stained and IHC-stained pathological sections from the tumor tissue extracted from the patient's breast. Then we used the IHC-stained WSI as the fixed image to register the H\&E-stained WSI: projection transformation and elastix registration enabled the H\&E-stained WSI to be globally and locally aligned with the IHC-stained WSI. Finally, we refined the transformed WSI and cut the H\&E-IHC WSI pair to obtain paired H\&E-IHC image patches.}
\label{pipeline}
\end{figure*}

\subsection{{Data Construction Process}}

As denoted in Fig.~\ref{pipeline}, the data construction process mainly includes the following steps: slice preparation, scanning, projection transformation, elastix registration, image refinement and patch selection.

During the preparation of a pathological slice pair, two layers of sections needed to be continuously cut out from the same tumor tissue for H\&E staining and IHC staining, respectively. Therefore, the section for H\&E staining was similar in shape to the corresponding section for IHC staining. The prepared pathological slides were then scanned into WSIs. Specifically, we used Hamamatsu NanoZommer S60 (capable of 20$\times$ magnification) to scan H\&E-stained WSI and the corresponding IHC-stained WSI. Due to computing power and memory limitations, we downsampled the original images by reducing their dimensions to half in both width and height in the subsequent processing steps.

Then the downsampled H\&E-IHC WSI pairs were aligned through image registration with two operations: projective transformation and elastix registration. The two transformation steps were employed to achieve alignment between the H\&E-stained WSI and corresponding IHC-stained WSI in terms of global contour and internal details, respectively.

\subsection{Image Registration}
To achieve the alignment of H\&E-stained and IHC-stained images, we implemented the following two-step registration process:

\textbf{Projective Transformation.}
First, we took the IHC-stained image as a reference and performed a projective transformation on the corresponding H\&E-stained image. Projective transformation requires no less than 4 selected one-to-one correspondences in the image to be transformed (H\&E) and the reference image (IHC). Projective transformation can then perform operations such as translation, scaling, rotation, beveling, and perspective distortion on the image to be transformed so that the H\&E-stained image and the IHC-stained image can be initially aligned on the outline.

\textbf{Elastix Registration.}
After completing the projective transformation of the H\&E-stained image, there were still some misalignments between the H\&E and IHC-stained images. We therefore iteratively registered H\&E and IHC-stained images using the medical image registration software elastix to further remove these misalignments. In the specific implementation process, since the computer memory cannot carry the high-resolution WSI registration calculation, we divided the H\&E and IHC-stained images into 16 parts for registration respectively, and then the registered parts were spliced back into WSI according to the original positional relationship.

\begin{figure}[!t]
\centering
\includegraphics[width=0.99\linewidth]{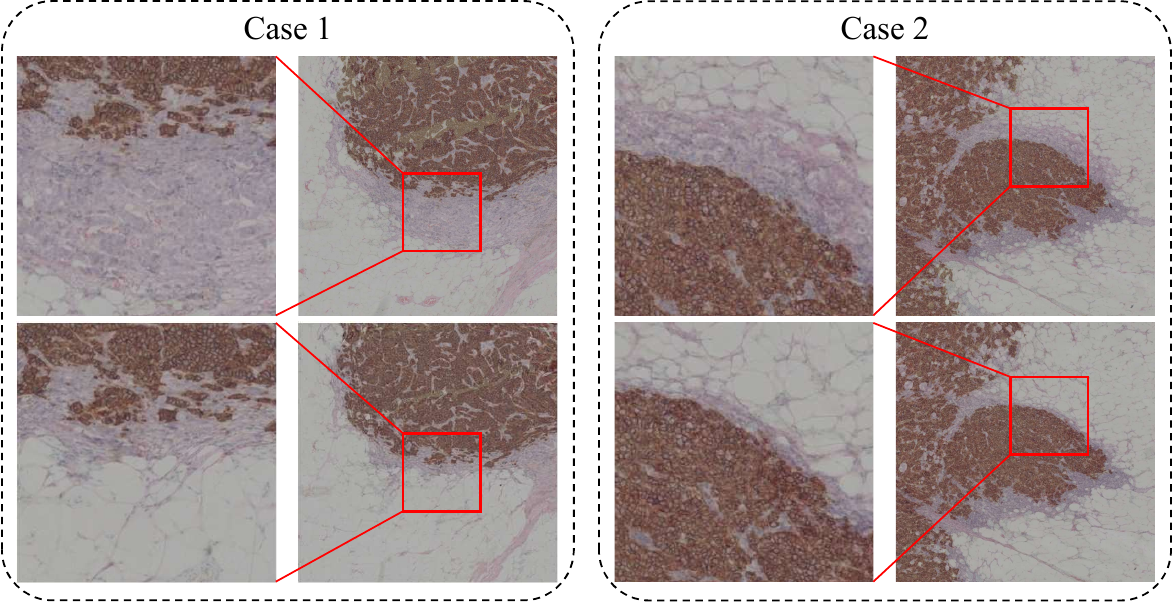}
\caption{Visual comparison of projection transformation results (the top row) and elastix registration results (the bottom row). By overlapping the transformed H\&E image with the corresponding IHC-stained image, it can be seen that elastix registration can improve the coincidence of the two images.}
\label{registration_compare}
\end{figure}

The visualization results of the projection transformation and elastix registration are compared in Fig.~\ref{registration_compare}.

\subsection{Post-Processing}

During the registration process, some operations in projective transformation (e.g. the rotation and scaling of the image) left some black areas at the edges of the WSI. At the same time, in the process of elastix registration, in order to align the internal content of the 16 image parts, the edges also moved to their inside resulting in black borders. We implemented an image refinement to address the above problems by filling the black area with surrounding pixels.
Finally, we segmented the WSIs into square patches with a side length of 1024 pixels and filtered out regions that did not contain tumor tissue or that were not aligned by the two-step registration procedure.
\vspace{0.2cm}

\section{Challenge Setup}

\subsection{Aims and Tasks}
This challenge aims to advance research on generating immunohistochemical images of breast cancer. The generated IHC-stained images should contain accurate HER2 expression information for direct interpretation by physicians. The use of a deep learning model for generating immunohistochemical images of breast cancer has the potential to save time, manpower, and material resources by eliminating the need for IHC-stained section preparation. Participants need to use pairs of H\&E and IHC-stained images in the training set to train an image translation model. During the prediction stage, the model must generate IHC-stained images using only H\&E-stained images as input. Furthermore, we annotated the extracted image patches using the HER2 expression information (0/1+/2+/3+) at the WSI level, which has been interpreted by pathologists. This label information is only available for model training and not for model prediction. Using label information is optional, but if a participant uses label information in their method, they should indicate it in the comments when submitting their results.

\subsection{Data Information}

\begin{figure}[!t]
\centering
\includegraphics[width=0.50\linewidth]{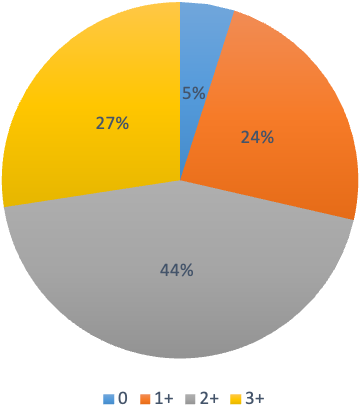}
\caption{The distribution of HER2 expression levels within the dataset.}
\label{level_distribution}
\end{figure}

Our dataset contains 4872 pairs of aligned H\&E-IHC pathology image patches, which come from the WSIs of more than 300 patients. The HER2 expression levels of these patients encompass four grades: 0, 1+, 2+, and 3+. Fig.~\ref{level_distribution} illustrates the distribution of HER2 expression levels of the image pairs (We utilize the HER2 expression level of the WSI to represent that of the corresponding image pair). The dataset used in this challenge consists of 3396 pairs of the training set images, 500 pairs of the validation set images, and 977 pairs of test set images. The three subsets are obtained by randomly partitioning these 4,872 pairs of images. Among them, the H\&E-IHC image pairs in the training set and validation set are all open to the participants, and for the testing set, only the H\&E-stained images are open to the participants. The validation set images are only for participants to test and optimize the performance of their models, and can not be used for model training. The test set images are used for challenge evaluation and to get the final ranking.

\subsection{Evaluation}

We used Peak Signal to Noise Ratio (PSNR) and Structural Similarity (SSIM) as the metrics to evaluate the quality of the generated images. PSNR is based on the error between the corresponding pixels of two images and is the most widely used objective evaluation index.

\begin{equation}
           PSNR(x,y) = 10{\times}\log_{10}{\frac{{(2^8-1)}^2}{MSE} },
  \label{PSNR}
\end{equation}

\begin{equation}
           MSE(x,y) = \frac{1}{mn}  {\textstyle \sum_{i=0}^{m-1}\sum_{j=0}^{n-1}\left [  x\left (  i,j\right ) -y(i,j)\right ]^{2} },
  \label{MSE}
\end{equation}
where $x$ and $y$ denote the generated IHC-stained image and the corresponding ground truth, $m$ and $n$ denote the width and height of the image. 

However, the evaluation result of PSNR may be different from the evaluation result of the Human Visual System (HVS). Therefore, we also used SSIM, which comprehensively measures the differences in image brightness, contrast, and structure.

\begin{equation}
           SSIM(x,y) = \frac{(2\mu_x\mu_y+C_1)(2\sigma _x\sigma_y+C_2)}{(\mu_{x}^{2}+\mu_{y}^{2}+C_1)(\sigma_{x}^{2}\sigma_{y}^{2}+C_2)}, 
  \label{SSIM}
\end{equation}
where $\mu_x$ and $\sigma_x$ denote the mean and standard deviation of the generated IHC-stained image, $\mu_y$ and $\sigma_y$ denote the mean and standard deviation of the ground truth, $C_1$ and $C_2$ are constants.

The final ranking of the challenge was calculated by the weighted average of the participants' SSIM ranking and PSNR ranking:
\begin{equation}
           R_{Final} = 0.4 \times R_{PSNR} + 0.6 \times R_{SSIM},
  \label{Final ranking}
\end{equation}
where $R_{Final}$ determines the final ranking (smaller $R_{Final}$ means higher ranking), $R_{PSNR}$ denotes the rank of PSNR and $R_{SSIM}$ denotes the rank of SSIM.

\section{Methods}

We have summarized the method descriptions submitted by five top-ranked teams in Table~\ref{table1}. Among these teams, three employed fully supervised image translation models (arpitdec5, Just4Fun, vivek23), while the other two utilized weakly supervised image translation models (lifangda02, stan9). With regard to the utilization of supplementary information, teams Just4Fun and stan9 incorporated WSI-level category labels (0, 1+, 2+, 3+) during the training of their respective models.

\begin{table}[htb]
\scriptsize
\caption{Brief Comparison of Top-Ranked Participating Methods.} 
\begin{center}{
\begin{tabular}{ccccc}
\hline
Team                  & \makecell{Basic\\architecture}  & \makecell{HER2 expression\\level used?} &\makecell{Supervision}   \\ \hline
arpitdec5              & Pyramid Pix2pix   & \XSolidBrush  & Full supervision       \\
Just4Fun               & Self-developed    & \Checkmark    & Full supervision       \\ 
lifangda02             & CUT               & \XSolidBrush  & Weak supervision       \\ 
stan9                  & WeCREST           & \Checkmark    & Weak supervision       \\
vivek23                & Pix2pix           & \XSolidBrush  & Full supervision       \\    
\hline
\end{tabular}}
\end{center}
\label{table1}
\end{table}

\subsection{arptidec5:}

\begin{figure}[!t]
\centering
\includegraphics[width=0.99\linewidth]{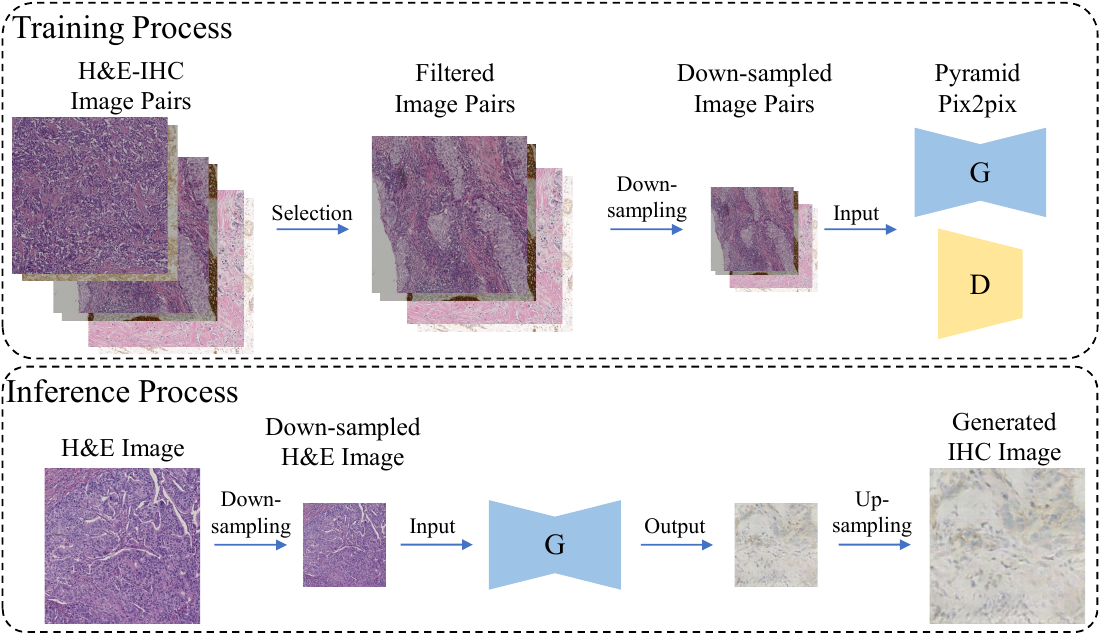}
\caption{During the training stage, the team arptidec5 used an open-source pathology image control software to filter out some images that were not suitable for training. Then they used downsampled H\&E-IHC image pairs to train a Pyramid Pix2pix model. During the inference stage, they made predictions based on the downsampled H\&E-stained images and then upsampled the generated IHC-stained images to the original resolution to obtain the final results.}
\label{arpit_method}
\end{figure}

The solution of team arptidec5 was built based on the framework of Pyramid Pix2pix in the BCI \cite{Liu_2022_CVPR}. In contrast to work \cite{Liu_2022_CVPR}, they filtered the image pairs in the dataset and performed downsampling during the data preprocessing stage. Specifically, before the images were used as input to the model for training, they went through a quality control process to ensure images with artifacts, cracked tissue, or blurriness were removed from the training process \cite{janowczyk2019histoqc}. Around 10\% of the images among the train set were not used for training. Then the images were resized from 1024$\times$1024 to 256$\times$256 and were used as input to the model. The output images obtained from the model were subsequently resized from 256$\times$256  to 1024$\times$1024 to match the original input dimensions. The training and inference process of team arptidec5 is illustrated in Fig.~\ref{arpit_method}.

\subsection{Just4Fun:}

Team Just4Fun proposed a level-aware BCIStainer to learn HER2 expression levels from H\&E-stained slices $I_{he}$, and meanwhile, translate $I_{he}$ to IHC-stained slices $\hat{I}_{ihc}$. The goals of BCIStainer are: (1) keeping consistency between HER2 expression levels of $\hat{I}_{ihc}$ and the ground truth $I_{ihc}$; (2) generating similar content of $\hat{I}_{ihc}$ as $I_{ihc}$; (3) making $\hat{I}_{ihc}$ and $I_{he}$ be structurally similar.

\begin{figure*}[!t]
\centering
\includegraphics[width=0.99\linewidth]{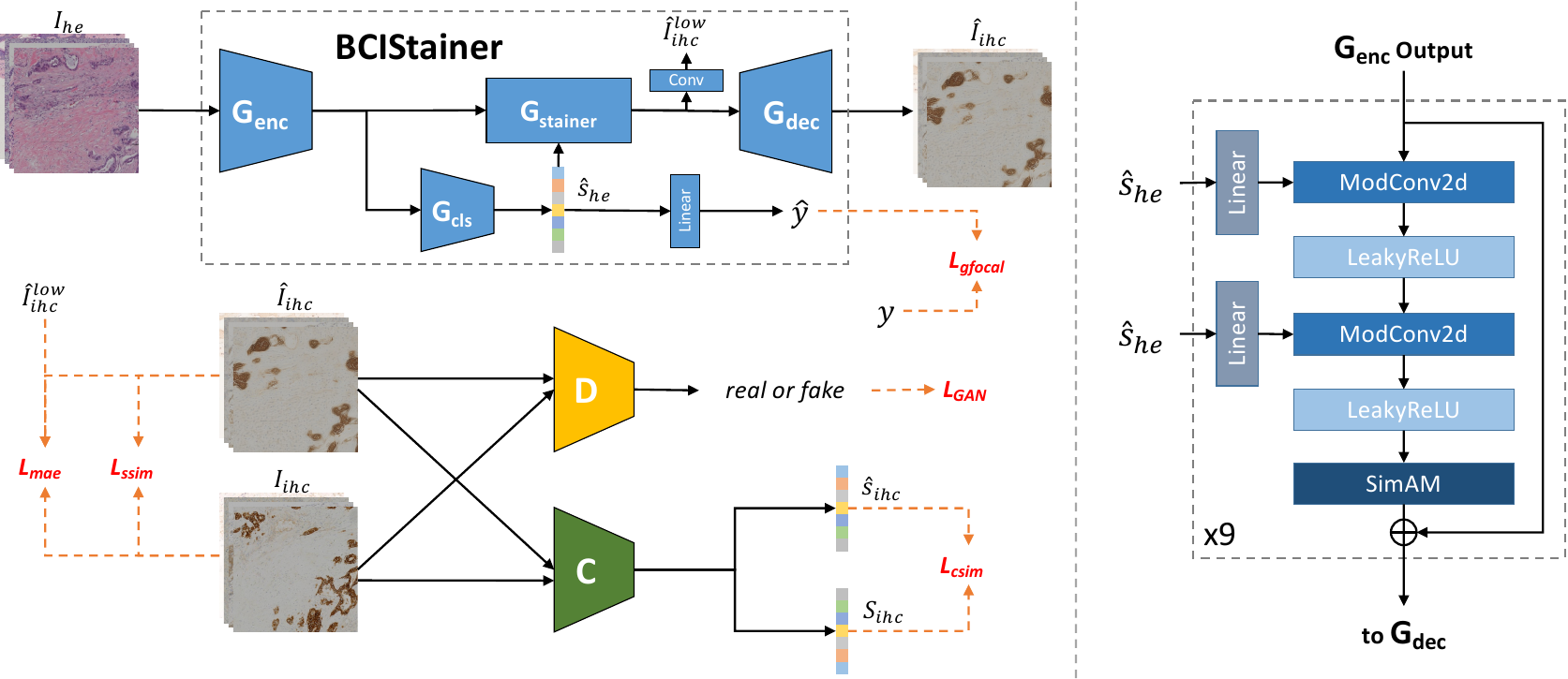}
\caption{Team Just4Fun used the architecture of generative adversarial network as a whole, in which the encoder of BCIStainer extracts features from the H\&E-stained image for classification, and the classification information, in turn, guides the generation of IHC-stained images. For the generated IHC-stained images, the participants used Mean Absolute Error (MAE), Structural Similarity (SSIM), and Cosine Similarity (CSIM) loss to constrain.}
\label{Just4Fun_method}
\end{figure*}

The architecture of HER2 expression level-aware BCIStainer $(G)$ is shown in Fig.~\ref{Just4Fun_method}. $G_{enc}$ encodes the input $I_{he}$ from high resolution (1024$\times$1024) to a feature map of size 128$\times$128. $G_{cls}$ outputs latent features $\hat{S}_{he}$ from encoded $I_{he}$, and then $\hat{S}_{he}$ is used for predicting HER2 expression $\hat{y}$ by a linear classification header. $G_{stainer}$ shown in Fig.~\ref{Just4Fun_method} applies $\hat{S}_{he}$ as the condition in the weight-demodulated layer \cite{karras2020analyzing} to perform the image-to-image translation process. The participants also added the parameters-free attention layer SimAM \cite{yang2021simam} in $G_{stainer}$ to enhance the salient features of cell structure or HER2 expressions, such as cell edges, cell nucleus, and dark regions of high-level HER2 expression. $G_{stainer}$ consists of 9 basic blocks. At the end of each basic block, they added a residual skip connection before the output. A convolutional layer is followed by $G_{stainer}$ and outputs the low-resolution prediction $\hat{I}_{ihc}^{low}$. The final $G_{dec}$ recovers translated feature maps to the original image space and outputs predicted IHC-stained slices $\hat{I}_{ihc}$.

The following is a detailed description of the loss functions used in this framework, including level loss, content loss, and adversarial loss.

\textbf{Level Loss.} Level loss is the combination of multiple-class focal loss $L_{gfocal}$ \cite{lin2017focal} and cosine similarity loss $L_{csim}$ in sample level. Because of the imbalanced multiple-class dataset, they used a $L_{gfocal}$ to train the classifier in BCIStainer with predicted level $\hat{y}$ and the ground truth level $y$. Thus $\hat{S}_{he}$ is able to conduct guidance to $G_{stainer}$. $L_{gfocal}$ can be calculated by the following equations:
\begin{equation}
           L_{gfocal} = \frac{1}{N} \sum_{n=1}^{N} -\alpha_n(1-p_{t,n})^{\gamma}log(p_{t,n}), 
  \label{L_gfocal}
\end{equation}

\begin{equation}
p_{t,n} =
    \begin{cases}
    p_n,  & \text{if $y=n$}, \\
    1-p_n, & \text{otherwise},
    \end{cases}
\label{ptn}
\end{equation}

\begin{equation}
p_n = \frac{e^{\hat{y}_n }}{\sum_{i=1}^{N}e^{\hat{y}_i} }, 
\label{pn}
\end{equation}
where $N$ is the number of expression levels, $n$ is the $n$th level, $i$ is the $i$th level, $p_n\in [0,1]$ is the predicted probability for the class with label $n$, $\alpha $ is a vector of weight coefficients of all levels, $(1-p_{t,n})^\gamma$ is the modulating factor to reshape the loss function to down-weight easy examples, and $\gamma$ is a tunable parameter to smoothly adjusts the modulating factor \cite{lin2017focal}.
To compute $L_{csim}$, they firstly trained a classifier as comparator $C$, only using the real IHC-stained slices $I_{ihc}$ and expression levels $y$ with a multiple-class focal loss $L_{cfocal}$ in the same formulation as \eqref{L_gfocal} to \eqref{pn}. $C$ generates representations $\hat{S}_{ihc}$ and $S_{ihc}$ by inputting $\hat{I}_{ihc}$ and $I_{ihc}$. $L_{csim}$ measures the similarity between $\hat{I}_{ihc}$ and $I_{ihc}$ in sample level as \eqref{L_csim}. Lower $L_{csim}$ reveals higher consistency in expression levels of $\hat{I}_{ihc}$ and $I_{ihc}$.
\begin{equation}
           L_{csim} = 1-\frac{\hat{S}_{ihc}·S_{ihc}}{\left \| \hat{S}_{ihc} \right \| ·\left \| S_{ihc} \right \| }.
  \label{L_csim}
\end{equation}

\textbf{Content Loss.} Content loss is composed of Mean Absolute Error (MAE) loss $L_{mae}$ and structure similarity loss $L_{ssim}$. $L_{mae}$ is used to measure content consistency between full resolution $\hat{I}_{ihc}$ and $I_{ihc}$, and also in the low resolution $\hat{I}_{ihc}^{low}$ and $I_{ihc}^{low}$. $L_{mae}$ is shown as \eqref{L_mae}:
\begin{equation}
           L_{mae} = \left \| I_{ihc}-\hat{I}_{ihc} \right \|_1+\left \| I_{ihc}^{low} - \hat{I}_{ihc}^{low} \right \|.
  \label{L_mae}
\end{equation}
Structure similarity comprehensively measures the differences between images in brightness, contrast, and structure. They applied structure similarity as a loss function to train the generator directly as \eqref{L_ssim}:
\begin{equation}
           L_{ssim} = 1-SSIM(\hat{I}_{ihc},I_{ihc}).
  \label{L_ssim}
\end{equation}
% where SSIM is implemented as \cite{zhao2016loss}.

\textbf{Adversarial Loss.} Adversarial loss $L_{GAN}$ is a multiple-scale version of PatchGan from Pix2pixHD \cite{wang2018high}. $L_{GAN}^{1024}$ is computed by $\hat{I}_{ihc}$ and $I_{ihc}$ in full resolution 1024$\times$1024. $L_{GAN}^{512}$ is computed in the same way as $L_{GAN}^{1024}$, but using resized $\hat{I}_{ihc}$ and $I_{ihc}$ in resolution 512$\times$512. $L_{GAN}$ is the mean of $L_{GAN}^{1024}$ and $L_{GAN}^{512}$ as \eqref{L_GAN}.

\begin{equation}
\begin{aligned}
           L_{GAN} &= arg\min_{G} \max_{D}(L_{GAN}^{1024}(\hat{I}_{ihc},I_{ihc})\\
           &+L_{GAN}^{512}(\hat{I}_{ihc},I_{ihc}))\times 0.5.
  \label{L_GAN}
  \end{aligned}
\end{equation}

\textbf{Overall Loss.} Overall loss is constructed as \eqref{L_overall}:
\begin{equation}
\begin{aligned}
           L &= \lambda_{gfocal}L_{gfocal}+\lambda _{csim}L_{csim}+\lambda _{mae}L_{mae} \\
           &+\lambda _{ssim}L_{ssim}+\lambda _{GAN}L_{GAN},
  \label{L_overall}
\end{aligned}
\end{equation}
where $\lambda_{gfocal}$, $\lambda_{csim}$, $\lambda_{mae}$, $\lambda_{ssim}$ and $\lambda_{GAN}$ are weighting-parameters for corresponding losses.

Team Just4Fun has open-sourced their code at github: \href{https://github.com/quqixun/BCIStainer}{https://github.com/quqixun/BCIStainer}.

\subsection{lifangda02}
\label{formats}
Team lifangda02 summarized the challenging aspects of H\&E-to-IHC translation into two points:

(1) Inconsistencies in the H\&E-IHC pairs. Since re-staining a slice is physically infeasible, a matching pair of H\&E-IHC slices are taken from two depth-wise consecutive cuts of the same tissue and scanned separately. This inevitably prevents pixel-perfect image correspondences due to morphology inconsistency and alignment error. The former is inherent to the fact that the image pair is from separate cuts and their preparation routines might differ. The latter is only exacerbated by the former in the image registration process.

(2) Reproducing the diagnosis-critical characteristics. In this challenge, IHC staining highlights tissue regions with a positive HER2 expression with a brownish color. The higher the HER2 expression is, the darker the brown and the higher the contrast against the benign tissue regions. Therefore, correctly reflecting the HER2 expression levels in the generated IHC-stained images is a huge challenge, especially given the much lower contrast levels between the malignant and benign regions in the H\&E-stained images. Additionally, doing so accurately and in a visually discriminative manner is of the core interest of H\&E-to-IHC translation.

To address the first challenge, the participants approached the problem of H\&E-to-IHC stain transfer from the perspective of “weakly” supervised image-to-image translation. The solution was built on top of the Contrastive Unpaired Translation (CUT) framework by \cite{park2020contrastive}. They augmented the CUT framework with a novel paired contrastive loss, aimed to mitigate the inconsistencies in the H\&E-IHC image pairs. To further partially address the second challenge, they designated the discriminator to classify the HER2 level as an auxiliary task.

The CUT framework ensures the content is consistent in the generated image by maximizing the mutual information between input and output. This is implemented by minimizing a patch-based InfoNCE contrastive loss, which aims to learn an embedding that associates corresponding patches to each other, while disassociating them from others. Given a query (a patch) in the output image, the positive is the corresponding patch and the negatives are noncorresponding patches, both from the input image. This loss is denoted as $L_{NCE}$. For more details, please refer to (3) in work \cite{park2020contrastive}.

The innovative contribution of team lifangda02 is the introduction of paired InfoNCE contrastive loss $L_{pNCE}$, which extends $L_{NCE}$ to paired images especially to combat the inconsistencies in H\&E-IHC image pairs. More specifically, given an output patch as query, the corresponding IHC-stained patch is designated as the positive and the noncorresponding patches are designated as the negatives. Then they used the same InfoNCE-based formulation for $L_{pNCE}$.

The key intuition behind $L_{pNCE}$ is that it can be seen as a soft image reconstruction learning criteria. Instead of using a predefined loss term that may not work well on inconsistent ground truth pairs, $L_{pNCE}$ punishes dissimilarities between the query and the positive in a learned latent space. 
Therefore, owing to this adaptiveness, $L_{pNCE}$ is more robust towards noisy supervision.

The participants used the resnet-9-blocks generator architecture with the loss function in \eqref{G*}:

\begin{equation}
\begin{aligned}
        G^{*} &= arg\min_{G} \max_{D} L_{GAN}+10{\times}L_{NCE}+10{\times}L_{pNCE} \\
        &+2{\times}L_{dis-cls}+20{\times}L_{multi-scale},
  \label{G*}
\end{aligned}
\end{equation}
where $L_{dis-cls}$ is the auxiliary cross-entropy-based classification loss using the discriminator and $L_{multi-scale}$ is the multi-scale image reconstruction loss as introduced in work \cite{Liu_2022_CVPR}.

An extended version of the method by team lifangda02 can be found in \cite{li2023adaptive}, where the authors further extended the $L_{pNCE}$ loss to adaptively learn from H\&E-IHC image pairs that are more consistent.

Team lifangda02 has open-sourced their code at github: \href{https://github.com/lifangda01/AdaptiveSupervisedPatchNCE}{https://github.com/lifangda01/AdaptiveSupervisedPatchNCE}.

\subsection{stan9:}

Supervised methods are generally the best methods if paired datasets are available in image translation. However, it is difficult to obtain well paired H\&E-IHC images. Unsupervised methods can carry out image translation using unpaired datasets, while the results are less than satisfactory. Therefore, weakly supervised learning may be a better way for H\&E to IHC-stained image translation. The method submitted by team stan9 is based on WeCREST \cite{dai2022weakly}, which is a weakly supervised deep generative network for style transformation. The backbone of their method is consistent with U-GAT-IT \cite{kim2019u}. In each iteration of the training, the input images were sampled according to the sampling rule, which is determined by \eqref{Qi}:
\begin{equation}
           Qi = \frac{1+\frac{cov(S_i,T_i)}{\sigma_{S_i}\sigma_{T_i}} }{\sqrt{ {\textstyle \sum_{j=1}^{N}}H_i·H_j} },
  \label{Qi}
\end{equation}
where $\sigma_{S_i}$ and $\sigma_{S_i}$ are the standard deviations of the source image $S_i$ and target image $T_i$, $cov(\cdot)$ is the covariance, $H_i$ is a vector of the normalized image histogram for image $i$, and $N$ is the total number of images. The discriminator can classify images into $N+1$ classes, including $N$ classes in the pool of real image styles and a fake image style.
In addition, the existing methods only consider style transformation but ignore the positive/negative consistency. It is hard for the network to identify cancer areas and the colors of generated images are incorrect sometimes. Therefore, the participants added an auxiliary classifier to the discriminator. The role of the classifier is to classify images according to the HER2 expression status. H\&E-stained images are labeled according to the expression status of the corresponding IHC-stained images. As the training goes on, the classification module can make the generator learn pathological characteristics and keep the expression status of the input image and the generated image consistent.
The loss function \eqref{L_G} of the generator is composed of adversarial loss $L_{adv}$, class loss $L_{class}$, cam loss $L_{cam}$ \cite{kim2019u}, and cycle loss $L_{cycle}$. 
\begin{equation}
           L_G = \lambda _{G1}  L_{adv}+\lambda_{G2}L_{class}+\lambda_{G3}L_{cam}+\lambda_{G4}L_{cycle},
  \label{L_G}
\end{equation}
where $\lambda _{G1}$, $\lambda _{G2}$, $\lambda _{G3}$, $\lambda _{G4}$ are weighting-parameters; adversarial loss $L_{adv}$ has two formats: $L_{adv}^{s\to t}$ and $L_{adv}^{t\to s}$ which are determined by \eqref{L_adv_s_t} and ~\ref{L_adv_t_s}, respectively.
\begin{equation}
           L_{adv}^{s\to t} = -\sum_{i}Y_i\ln_{}{(D_t(T_i))}+Y_0\ln_{}{(D_t(G_{s\to t}(S_i)))},
  \label{L_adv_s_t}
\end{equation}
where $D_t$ is the discriminator of HER2 images, $Y_i$ and $Y_0$ are one-hot style labels for image $i$ and the fake images.
\begin{equation}
           L_{adv}^{t\to s} = -\sum_{i}Y_i\ln_{}{(D_s(S_i))}+Y_0\ln_{}{(D_s(G_{t\to s}(T_i)))},
  \label{L_adv_t_s}
\end{equation}
where $D_s$ is the discriminator of H\&E-stained images, $Y_i$ and $Y_0$ are one-hot style labels for image $i$ and the fake images. The loss function of the discriminator (\eqref{L_D}) is composed of adversarial loss $L_{adv}$, class loss $L_{class}$, and cam loss $L_{cam}$. 

\begin{equation}
           L_D = \lambda_{D1}L_{adv}+\lambda_{D2}L_{class}+\lambda_{D3}L_{cam},
  \label{L_D}
\end{equation}
where $\lambda _{D1}$, $\lambda _{D2}$, $\lambda _{D3}$ are weighting-parameters.

\begin{figure}[!t]
\centering
\includegraphics[width=0.99\linewidth]{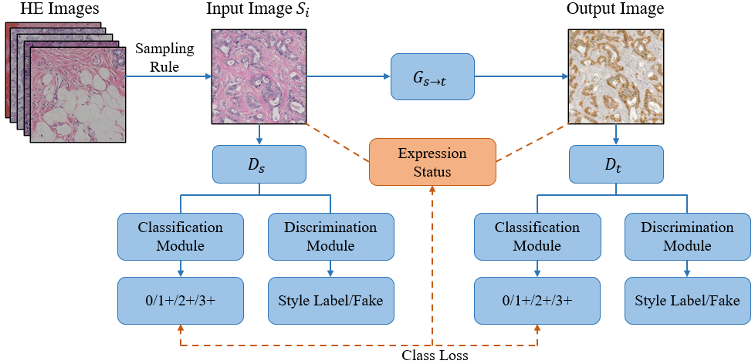}
\caption{Team stan9 used a dual image translation model. The discriminators of the two branches classify HER2 expression levels while not only distinguishing whether the images are real or fake. The classification module can make the generator learn pathological characteristics and keep the expression status of the input image and the generated image consistent.}
\label{stan9_method}
\end{figure}

\subsection{vivek23:}
\begin{figure}[!t]
\centering
\includegraphics[width=0.99\linewidth]{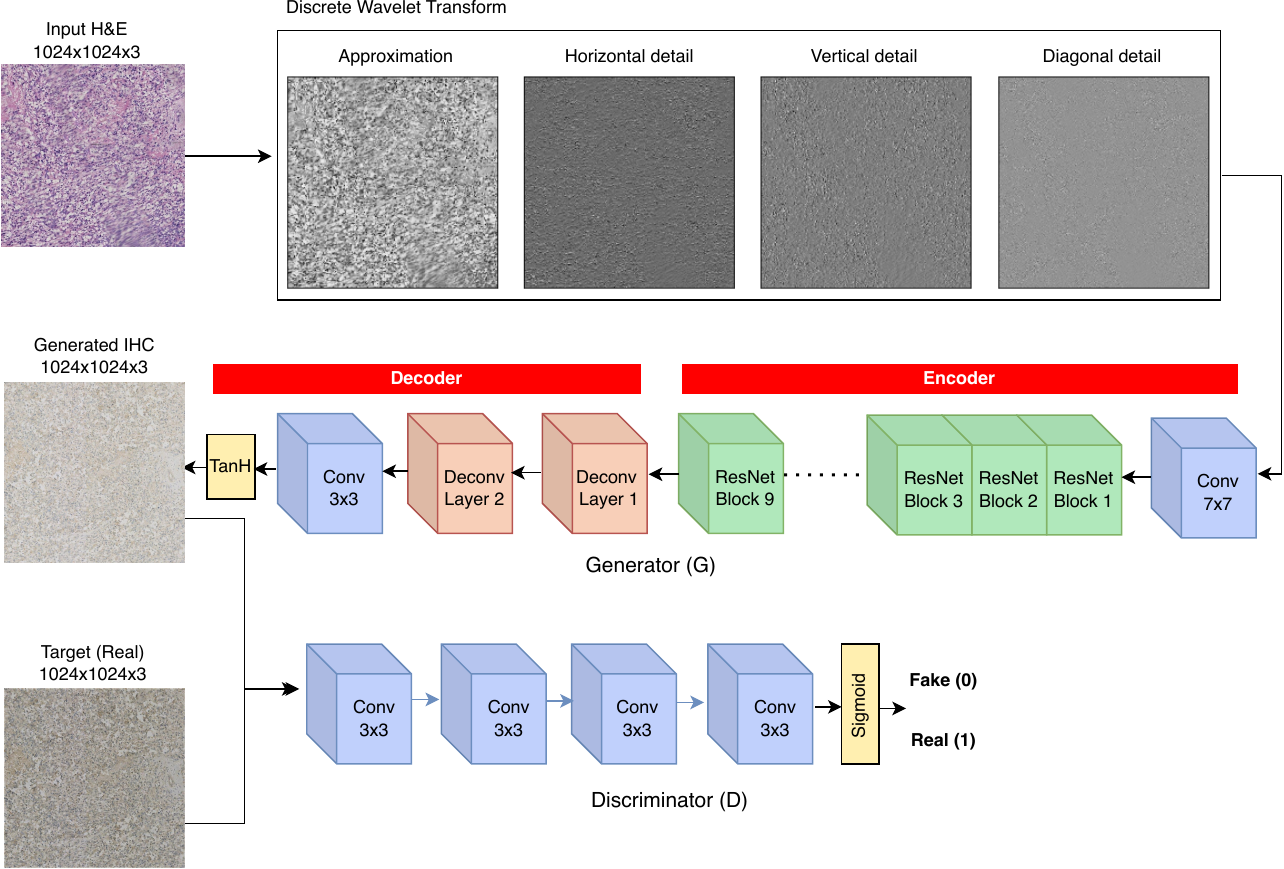}
\caption{Team vivek23 used Discrete Wavelet Transform (DWT) to convert RGB images into 4-channel spatial and frequency domain information, then input them into the network. DWT segments low-frequency and high-frequency details in an image, helping to accurately measure sharp changes in the input image.}
\label{vivek23_method}
\end{figure}

Fig.~\ref{vivek23_method} presents a general overview of the proposed wavelet-based pix2pix model that generates an IHC-stained image from the H\&E-stained source image. Team vived23 employed conditional generative adversarial network (cGAN) \cite{isola2017image} based on the paired images. The model comprises two sub-networks: a generator that generates a fake or synthetic image, and a discriminator that classifies the generated (fake) image against the corresponding ground truth. The generator network consists of an encoder and a decoder. In the encoder, they incorporated 9 intermediate residual blocks from the ImageNet pre-trained ResNet18 network \cite{he2016deep}.  Unlike traditional color image-based encoder that works directly on three color channels, they applied the discrete wavelet transform (DWT) to extract four-channel spatial and frequency domain features from the given H\&E images \cite{singh2022prior}. DWT splits the lower and higher-frequency details into sub-bands that help in precisely measuring sharp changes in the input image. In the generator network's architecture, the first convolutional layer employs 64 filters with a kernel size of 7$\times$7 and stride 1, followed by the InstanceNorm and the ReLU activation function \cite{zhu2017unpaired}. While the encoded features are decoded by the two ConvTranspose2D deconvolutional layers.

This team defined $i$ as the input H\&E image, $gt$ as the corresponding IHC-stained image, and $z$ as a random variable. Generator $G$ and Discriminator $D$, generate the output of $G (i, z)$ and $D (i, G (i, z))$, respectively. Therefore, the loss function of the generator network $G$ consists of the Binary CrossEntropy (BCE) loss and $L1$ loss that can be formulated as \eqref{L_gen}:
\begin{equation}
\begin{aligned}
           L_{Gen}(G,D) &= E_{i,gt,z}(-\log_{}{D(i,G(i,z))} ) \\
           &+\lambda E_{i,gt,z}(L_1(gt,G(i,z))),
  \label{L_gen}
  \end{aligned}
\end{equation}
where $\lambda$ is an empirical weighting factor that is set to 100. The $L1$ loss helps in reducing the number of false positives and fosters the training process by generating sharp images.
The discriminator network utilizes five convolutional layers with a kernel of 4$\times$4, and stride of 2$\times$2. The early four layers are followed by batch normalization and non-linear leaky ReLU (slope 0.2) activation function. The last layer uses a sigmoid activation function to discriminate between real and fake generated outcomes. The discriminator loss can be defined as \eqref{L_dis}:
\begin{equation}
\begin{aligned}
    L_{Dis}(G,D) &= E_{i,gt,z}(-\log_{}{(D(i,gt))} ) \\
           &+\lambda E_{i,gt,z}(-\log_{}{(1-D(i,G(i,z)))}).
  \label{L_dis}
\end{aligned}
\end{equation}     
Equation \eqref{L_dis} utilizes the BCE loss function for real IHC-stained images against the generated (fake) ones. During training, the discriminator network enforces the generator network in generating a better synthetic IHC-stained image while comparing it with actual ground truth. The discriminator network is not involved in the model’s evaluation phase.

\section{Results and Discussion}

\begin{figure}[!t]
\centering
\includegraphics[width=0.99\linewidth]{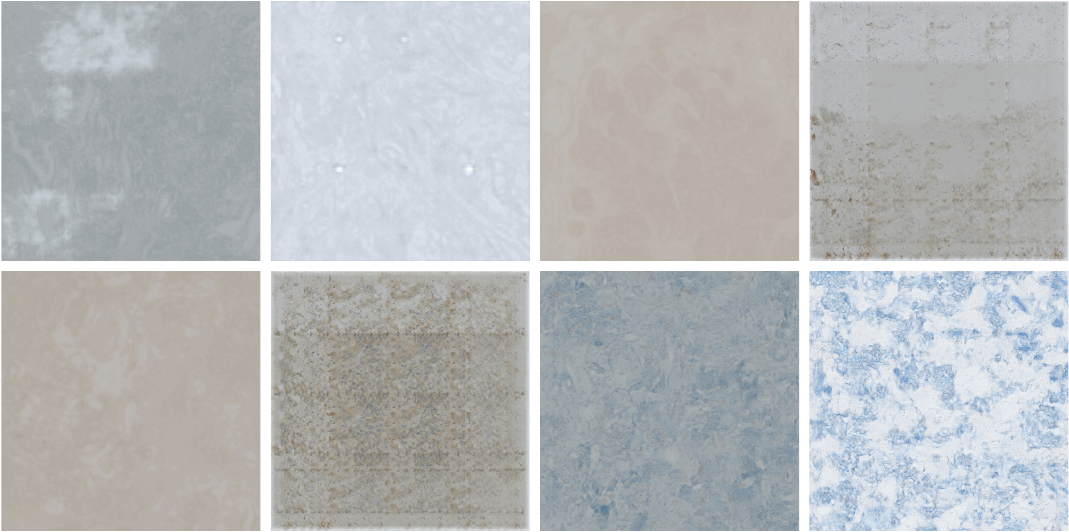}
\caption{Visualization of invalid submissions. These images are visually blurry. Pathologists cannot give an interpretation of HER2 expression levels based on these images.}
\label{failure_submission}
\end{figure}

We received submissions from a total of 12 teams for our challenge. To evaluate the generated immunohistochemical images submitted by the participants, we invited three doctors to review them. During the review process, we noticed that some of the participants' submissions contained blurry images where the cellular structure could not be observed, as shown in Fig.~\ref{failure_submission}. Due to their lack of clinical reference value, these submissions were deemed invalid. 

Ultimately, we were able to collect 6 valid submissions, and only these submissions were ranked.

\subsection{Quantitative Results}

Table~\ref{metrics} shows the PSNR and SSIM metrics of the participants in the challenge. Among them, the team arpitdec5 achieved the highest SSIM and the second highest PSNR, and according to \eqref{Final ranking}, they obtained the highest final ranking. The team Just4Fun obtained the highest PSNR and the second highest SSIM and finally ranked second. Finally, the third to sixth rankings were awarded to teams lifangda02, stan9, guanxianchao, and vivek23, respectively.

\begin{table}[htb]
\caption{Quantitative results of participating teams.} 
\begin{center}{
\begin{threeparttable}
\begin{tabular}{cccc}
\hline
Team           &Final Rank        & PSNR(dB)/Rank  & SSIM/Rank  \\ \hline
arpitdec5      &1        & 19.736 / 2       & 0.574 / 1   \\
Just4Fun       &2        & 22.929 / 1       & 0.559 / 2   \\ 
lifangda02     &3        & 17.927 / 5       & 0.555 / 3   \\ 
stan9          &4        & 17.959 / 4       & 0.543 / 4   \\
guanxianchao\tnote{1}   &5        & 19.560 / 3       & 0.497 / 5   \\
vivek23        &6        & 15.271 / 6       & 0.493 / 6   \\    
\hline
\end{tabular}

\label{metrics}
\begin{tablenotes}
\footnotesize
\item[1] Team guanxianchao didn't submit their method.
\end{tablenotes}
\end{threeparttable}}
\end{center}
\end{table}

\begin{figure*}[htbp]
\centering
    \begin{minipage}[t]{0.92\linewidth}
        \centering
        \includegraphics[width=\textwidth]{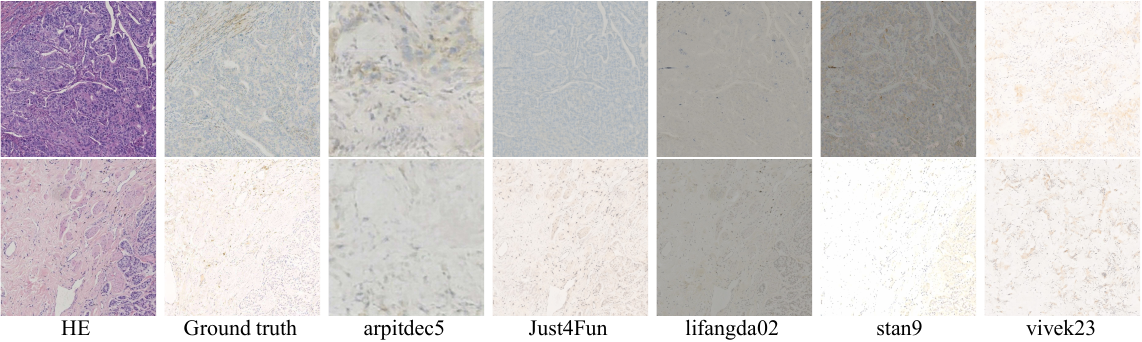}
        \centerline{\footnotesize{(a) IHC 0}}
        \vspace{0.05cm}
    \end{minipage}
    \begin{minipage}[t]{0.92\linewidth}
        \centering
        \includegraphics[width=\textwidth]{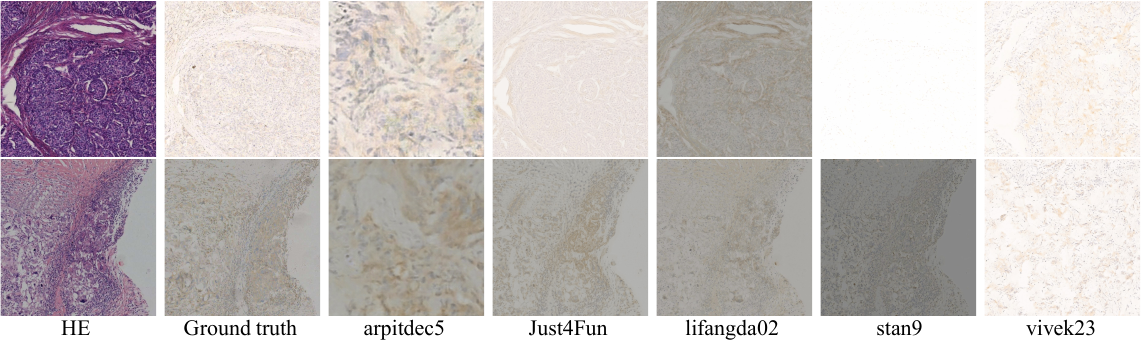}
        \centerline{\footnotesize{(b) IHC 1+}}
        \vspace{0.05cm}
    \end{minipage}
    \begin{minipage}[t]{0.92\linewidth}
        \centering
        \includegraphics[width=\textwidth]{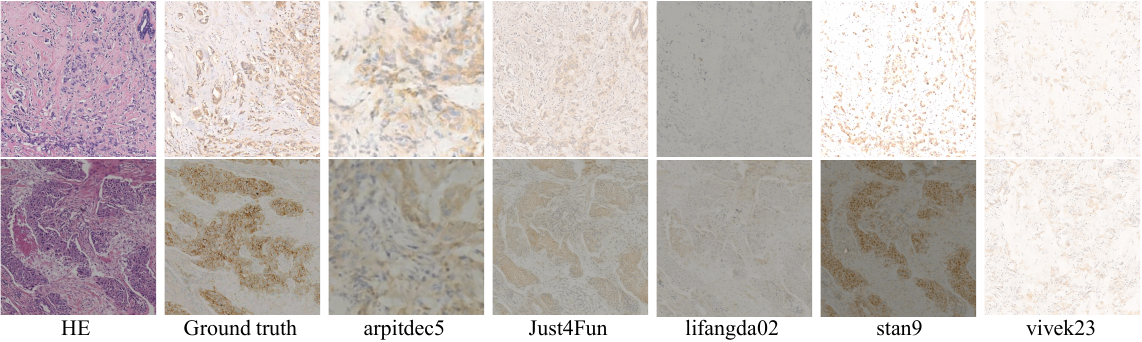}
        \centerline{\footnotesize{(c) IHC 2+}}
        \vspace{0.05cm}
    \end{minipage}
    \begin{minipage}[t]{0.92\linewidth}
        \centering
        \includegraphics[width=\textwidth]{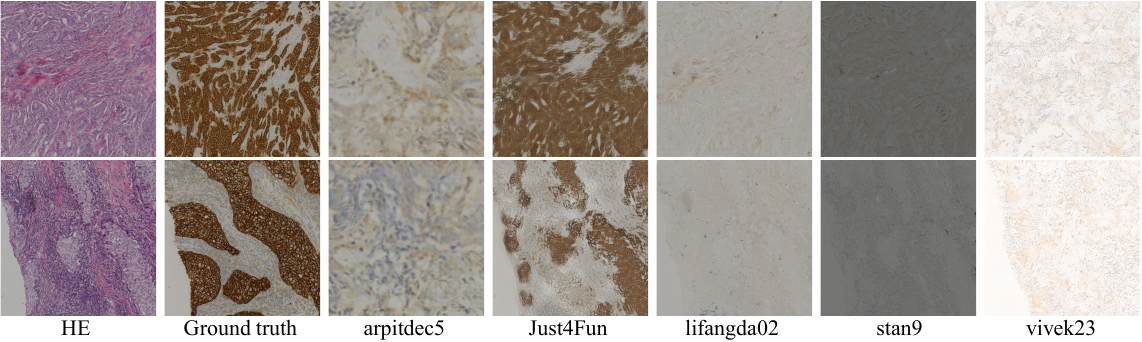}
        \centerline{\footnotesize{(d) IHC 3+}}
    \end{minipage}
    \caption{Visualization results submitted by participating teams. Subfigures (a)-(d) show the generated IHC-stained images at different HER2 expression levels.}
    \label{results}
\end{figure*}

\subsection{Qualitative Results}
Fig.~\ref{results} shows the generated IHC-stained images submitted by the participants. In the results submitted by the team Just4Fun, the color depth of the cells is overall consistent with the ground truth, which indicates that their results can accurately reflect the expression level of HER2 to a large extent. The results submitted by the team stan9 can approach the ground truth when HER2 expression levels are 0, 1+, and 2+, but when HER2 is highly expressed (3+), their results cannot accurately reflect the HER2 expression level. The results of the other four teams show similar staining intensity at HER2 expression levels of 0, 1+, 2+, and 3+, and these results cannot accurately reflect the information on HER2 expression levels.

\subsection{Discussion}

The difficulty of this challenge is how to generate IHC-stained images that can correctly reflect the expression level of HER2 based on H\&E images. Most of the participants' methods have the problem of not being able to identify the high-level expression of HER2 in breast tissue. For example, the coloring degree of the IHC-stained images generated by the teams arpitdec5, lifangda02 and vivek23 almost have the same tone, showing light brown, which is a phenomenon of mode collapse. The images submitted by Just4fun can reflect the expression level of HER2 to a large extent: when the expression level of HER2 is low (0/1+), the generated IHC-stained images are lightly colored, and when the expression level of HER2 is high (2+/3+), the generated IHC-stained images exhibit darker coloration.

The ability of the generated IHC-stained images to accurately reflect the HER2 expression level is highly related to whether the corresponding method uses HER2 expression level information. Team Just4fun made good use of the category information of the original WSIs: their method uses category information to supervise the classification of H\&E-stained images ($G_{class}$), and uses H\&E category information to intervene in the generation of IHC-stained images; at the same time, the method also inputs the generated IHC-stained images and the real IHC-stained images into the classification network $C$, and uses cosine similarity loss to constrain the category information of the two to be consistent. The team stan9 also used label information. Their methed uses a cycle structure similar to cycleGAN~\cite{zhu2017unpaired}, and adds a classification branch to the two discriminators, constraining the two discriminators to agree on the classification of an H\&E-stained image and the corresponding IHC-stained image. Therefore, the results submitted by the team stan9 can also generate IHC-stained images with different degrees of coloring, and achieve relatively accurate results in the case of IHC 2+, but still cannot generate correctly stained IHC images in the case of IHC 3+.

\section{Conclusion}
The results of this breast cancer immunohistochemical image generation challenge indicate that producing high-quality IHC-stained images from H\&E-stained images, which accurately depict HER2 expression levels, remains a significant challenge. However, the methods submitted by the participants still provide many novel ideas for IHC-stained image generation.

% In this breast cancer immunohistochemical image generation challenge, the results demonstrate that generating high-quality IHC-stained images from H\&E-stained images that accurately reflect HER2 expression levels is still a significant challenge, however, the methods submitted by the participants still provide many novel ideas for IHC-stained image generation.

According to the collected methods, the integration of WSI-level label information can make the generated IHC-stained images avoid mode collapse and correctly reflect the expression level of HER2 to a certain extent. At the same time, the strongly constrained L1 loss used in the fully supervised methods may affect the quality of the generated image, while the weakly supervised or unsupervised methods do not impose pixel-level constraints on the images, so they can better maintain the cell structure in the pathological image. The weakly supervised/unsupervised image translation algorithm has great potential in the field of pathological image generation. In order to promote related research, we will release unaligned H\&E-IHC image pairs as an expansion of the BCI dataset. To explore more possibilities of breast cancer pathological image generation, we have opened the post-challenge submission phase, and scholars can still submit their results on Grand Challenge: \href{https://bci.grand-challenge.org}{https://bci.grand-challenge.org}.

\appendices

\section*{Acknowledgment}
Arpit Aggarwal and Anant Madabhushi are with Biomedical Engineering Department, Georgia Tech and Emory University (e-mail: aagga56@emory.edu, anantm@emory.edu). Anant Madabhushi is also a Research Career Scientist with the Atlanta Veterans Affairs Medical Center.

Germán Corredor is with Biomedical Engineering Department, Emory University (e-mail: gcorred@emory.edu).

Qixun Qu is an independent researcher (e-mail: quqixun@gmail.com).

Hongwei Fan is with Data Science Institute, Imperial College London (e-mail: h.fan21@imperial.ac.uk).

Fangda Li is with the School of Computer and Electrical Engineering, Purdue University (e-mail: li1208@purdue.edu).

Yueheng Li, Xianchao Guan and Yongbing Zhang are with the School of Computer Science and Technology, Harbin Institute of Technology (Shenzhen) (e-mail: 22s051028@stu.hit.edu.cn; 21s051007@stu.hit.edu.cn; ybzhang08@hit.edu.cn).

Vivek Kumar Singh is with the Department of Computer Engineering and Mathematics, Rovira I Virgili University (e-mail: vivekkr.singh90@gmail.com).

Farhan Akram is with the Department of Pathology and Clinical Bioinformatics, Erasmus Medical Center (e-mail: f.akram@erasmusmc.nl).

Md. Mostafa Kamal Sarker is with the Institute of Biomedical Engineering, University of Oxford, Oxford, UK (e-mail: md.sarker@eng.ox.ac.uk).

\bibliographystyle{ieeetr}
\bibliography{references}  

\end{document}